\documentclass{emulateapj}
\usepackage{amsmath}
\usepackage{graphicx}
\usepackage{amsfonts}
\usepackage{natbib}
\usepackage{color}
\usepackage{epstopdf}
\usepackage{graphicx}
\usepackage{multirow}

\begin{document}

\title[$\gamma$-ray QPO in PKS 0301$-$243]{A $\gamma$-ray Quasi-Periodic modulation in the Blazar PKS 0301$-$243?}
\author{Peng-Fei~Zhang\altaffilmark{1,2}, Da-Hai~Yan\altaffilmark{3,4,5}, Jia-Neng~Zhou\altaffilmark{6}, Yi-Zhong~Fan\altaffilmark{1}, Jian-Cheng~Wang\altaffilmark{3,4,5}, and Li~Zhang\altaffilmark{2}}
\altaffiltext{1}{Key Laboratory of Dark Matter and Space Astronomy, Purple Mountain Observatory, Chinese Academy of Sciences, Nanjing 210008, China; zhangpengfee@pmo.ac.cn; yzfan@pmo.ac.cn}
\altaffiltext{2}{Key Laboratory of Astroparticle Physics of Yunnan Province, Yunnan University, Kunming 650091, China}
\altaffiltext{3}{Yunnan Observatory, Chinese Academy of Sciences, Kunming 650011, China; yandahai@ynao.ac.cn}
\altaffiltext{4}{Center for Astronomical Mega-Science, Chinese Academy of Sciences, 20A Datun Road, Chaoyang District, Beijing, 100012, China}
\altaffiltext{5}{Key Laboratory for the Structure and Evolution of Celestial Objects, Chinese Academy of Sciences, Kunming 650011, China}
\altaffiltext{6}{Shanghai Astronomical Observatory, Chinese Academy of Sciences, 80 Nandan Road, Shanghai 200030, China}

\begin{abstract}

We report a nominally high-confidence $\gamma$-ray quasi-periodic modulation in the blazar PKS 0301$-$243. 
For this target, we analyze its \emph{Fermi}-LAT Pass 8 data covering from 2008 August to 2017 May.
Two techniques, i.e., the maximum likelihood optimization and the exposure-weighted aperture photometry,
 are used to build the $\gamma$-ray light curves. Then both the Lomb-Scargle Periodogram and the
Weighted Wavelet Z-transform are applied to the light curves to search for period signals.
A quasi-periodicity with a period of $2.1\pm0.3$ yr appears at the significance level of $\sim5\sigma$,
although it should be noted that this putative quasi-period variability is seen in a data set barely four times longer.
We speculate that this $\gamma$-ray quasi-periodic modulation might be evidence of a binary supermassive black hole.

\end{abstract}

\bigskip
\bigskip


\section{INTRODUCTION}
\label{sec:intro}

Blazars are a subclass of radio-loud active galactic nuclei (AGNs) whose relativistic jets almost point at observers \citep{Urry1995}.
It is generally thought that a central supermassive black hole (SMBH) provides the energy that powers the relativistic jet through BH spin or rotating accretion disk.
The emissions from blazar are dominated by the nonthermal emission from
the relativistic jet, extending from MHz radio frequencies to TeV $\gamma$-rays energies, also exhibiting variabilities at all energies
on a wide range of timescales.
The typical mutiwavelength
spectral energy distribution (SED) is distinguished by two broad peaks: a synchrotron component peaking at infrared to X-ray bands 
and a Compton component peaking in MeV to GeV energies.

The periodic variabilities of blazars have been extensively investigated in optical band
\citep[e.g.,][]{Bai1998,Bai1999,Fan2000,Xie2008,Li2009,Urry2011,King2013,Zhang2014,Bhatta2016,Fan16}.
An interesting case is OJ 287 with a $\sim$ 12-yr period cycle \citep{Kidger1992,Valtonen2006}. 
Searches for $\gamma$-ray quasi-periodic oscillations (QPOs) became possible after the launching of the \emph{Fermi Gamma-ray Space Telescope} in 2008.
So far, the Large Area Telescope \citep[LAT;][]{Abdo2009,Atwood2009} onboard \emph{ Fermi} has collected gamma rays over 8 yr.
The possible quasi-periodic variabilities of blazars with period of $\sim$2-3 yr have been
reported in $\gamma$-ray fluxes of several blazars \citep[e.g.,][]{Sandrinelli2014,1553,Sandrinelli2016a,Sandrinelli2016b,Sandrinelli2017,Zhang2017a,Zhang2017b}.
In particular, PKS 2155$-$304 have been reported having significance of $\sim$ 4$\sigma$ \citep{Zhang2017a}.
The quasi-periodic modulations in the blazars carry important information on the BH - jet system.

We present the results of searching for QPO in the $\gamma$-ray light curves of blazar PKS 0301$-$243.
A clear quasi-periodic signal with a period cycle of $\sim2.1$-yr  is found  at the significance level of $\sim$5$\sigma$,
though given that the full data set is only 8.78 years long this signal can easily have arisen randomly \citep[e.g. ][]{press}.
The paper is organized as follows: the detailed LAT data analysis and the main results are reported in Section 2.
In Section 3 we summary the results and present a brief discussion on the findings.


\section{Observations and Analysis}
\label{sec:Observations}

PKS 0301$-$243 is a high-synchrotron-peaked blazar (HSP) with its synchrotron peak frequency $\approx10^{15}\ $Hz  \citep{abr}, and its redshift is 0.266 \citep{pita}.
 The High Energy Stereoscopic System (H.E.S.S.) has detected TeV photons from this source \citep{abr}.

The events are collected between 2008 August 4 and 2017 May 19 (Modified Julian Date, MJD: 54,682.66 -- 57,892.66) in the energy range from 100 MeV to 500 GeV, 
and in a square region of interest (ROI) of $20^\circ\times20^\circ$
centered at the position of PKS 0301$-$243.
The position of the target is located at right ascension
(R.A.) = $\rm 03^h~03.442^m$, declination (decl.) = $\rm -24^h~07.192^m$ (J2000; $l=214.621,~b=-60.177$).
The analysis is performed with the \emph{Fermi} Science Tools version v10r0p5 package
which is provided in the Fermi Science Support Center (FSSC).\footnote{https://fermi.gsfc.nasa.gov/ssc/data/analysis/software/}
The Pass 8 LAT data \citep{Atwood2013} are used with keeping only the SOURCE class photon-like events
(with options evclass = 128 and evtype = 3 in the Tool $gtselect$). To minimize the contamination due to the gamma-rays bright Earth limb,
we exclude the events with zenith angles $\geqslant~90^{\circ}$. By running Tool $gtmktime$, we obtain the good time intervals
with high-quality photons.
All the data reductions follow the data analysis thread provided by FSSC
\footnote{https://fermi.gsfc.nasa.gov/ssc/data/analysis/scitools/}.
We adopt the instrumental response function (IRF) `P8R2\_SOURCE\_V6' in the analysis.
Two diffuse model files\footnote{https://fermi.gsfc.nasa.gov/ssc/data/access/lat/BackgroundModels.html}, namely gll\_iem\_v06.fit and iso\_P8R2\_SOURCE\_V6\_v06.txt, 
are used to model the Galactic and extragalactic diffuse $\gamma$-rays .
A binned maximum likelihood is adopted to fit the events in the whole time range with the model file generated with the script \emph{make3FGLxml.py}.
This file contains the information on the spectral parameters of all known 3FGL sources \citep{Acero2015} in the ROI.
The $\gamma$-ray spectrum of the target is power-law in the \emph{Fermi} 3FGL.
The best-fitting results are derived with \emph{Fermi} Tool \emph{gtlike}, and are saved as a new model file.
We also use the spectra in \emph{Fermi} 3FGL model file to fit the events in the square ROI.
The integrated photon flux of the best-fitting results above 100 MeV is $\rm F_{0.1-500~GeV}=(4.2\pm0.1)\times10^{-8}~photons\ cm^{-2}\ s^{-1}$,
and the index of power law is $1.90\pm0.01$ with the TS value of 9391.7 (the results in this paper with statistical errors only).
We construct the light curves based on this new model file.

\subsection{$\gamma$-ray light-curve}
\label{subsec:make lc} 
\begin{figure*}
\centering
\includegraphics[scale=0.5]{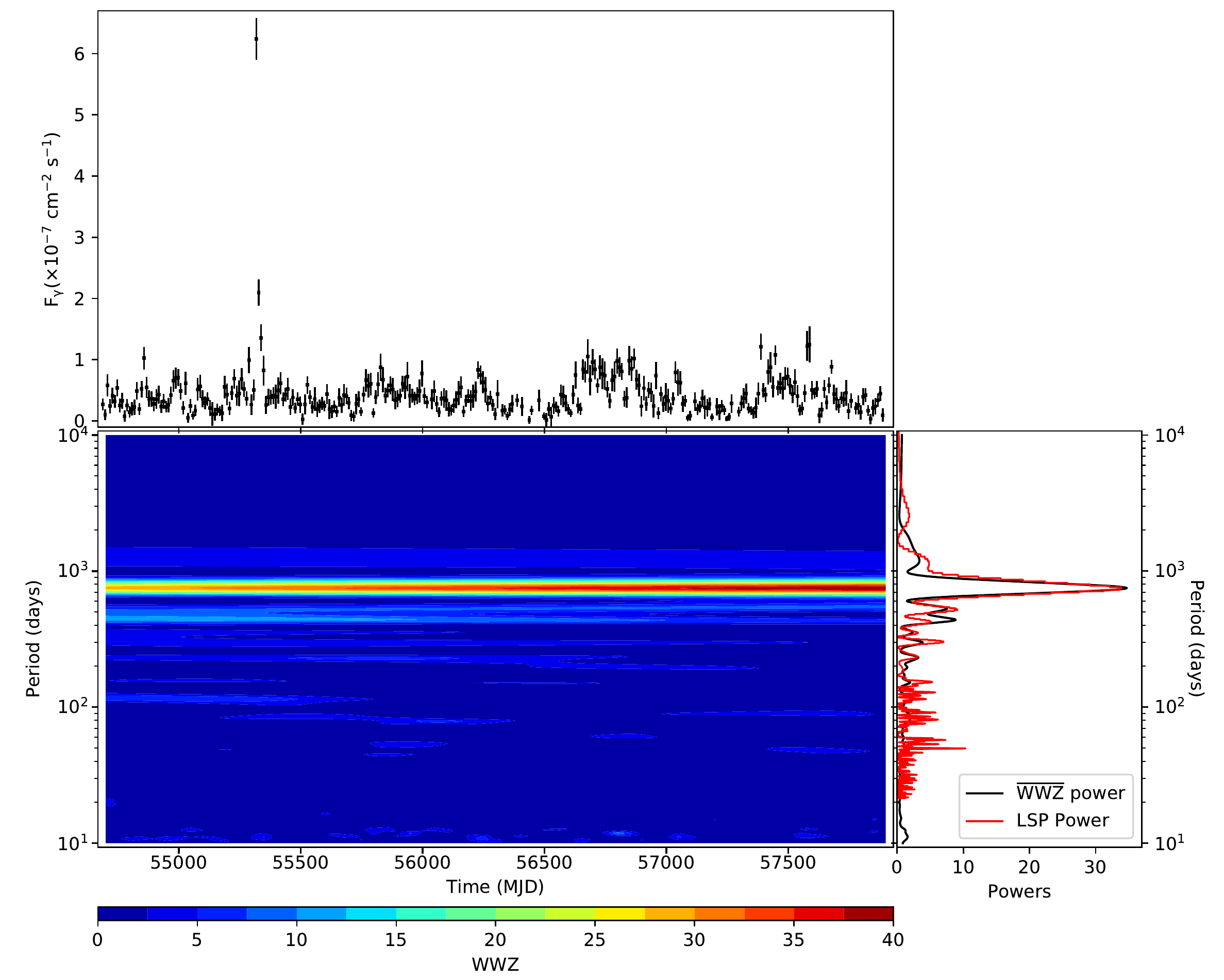}
\caption{Upper left panel: the 10-day-bin ML $\gamma$-ray light-curve.
		 Lower left panel: the 2D plane contour plot of the WWZ power of the light-curve.
		 Lower right panel: the LSP power spectrum for the light-curve (red solid line) and
		                    the time-averaged WWZ power (black solid line).}
\label{PKS_0301-243_10d}
\end{figure*}

\begin{figure*}
\centering
\includegraphics[scale=0.5]{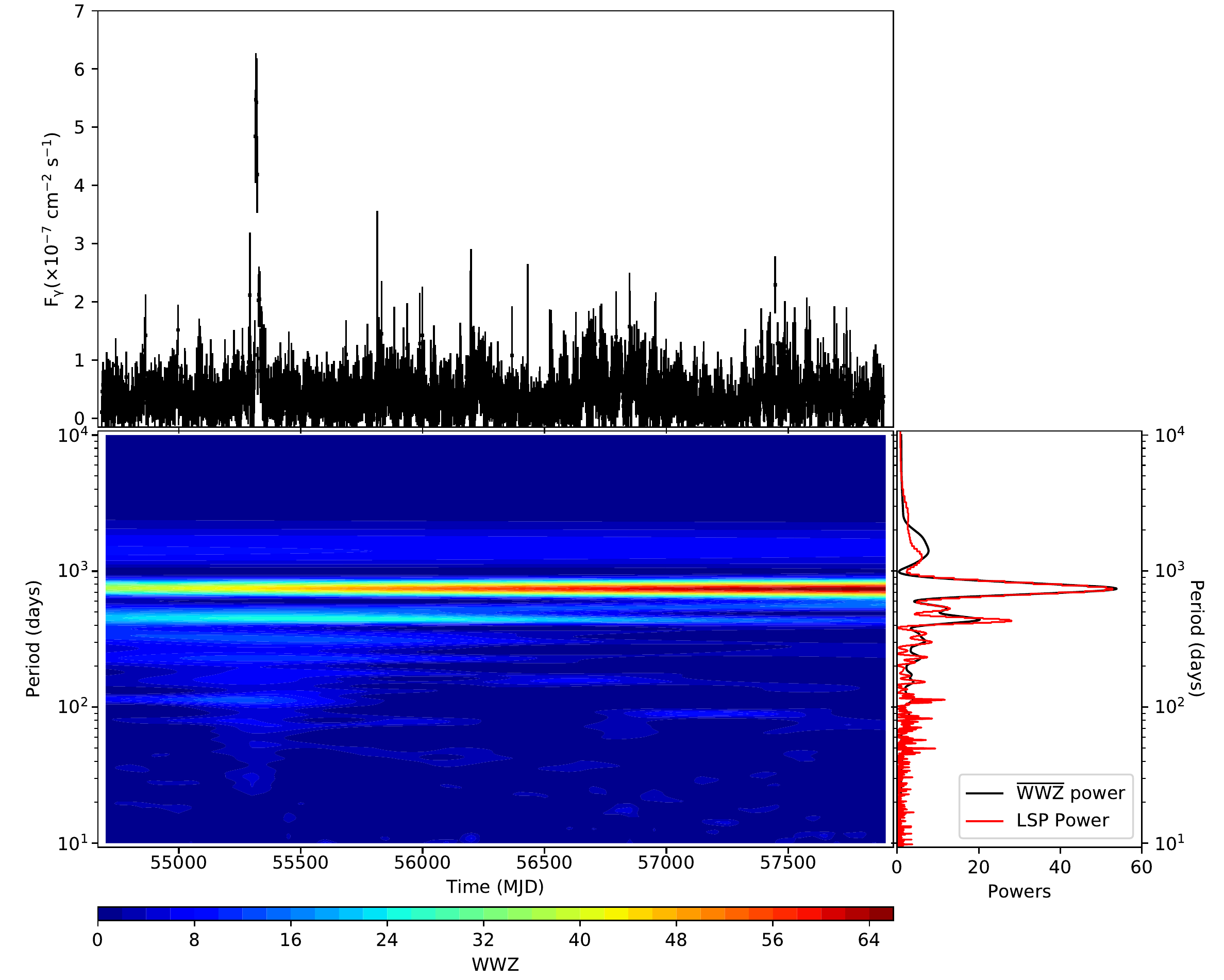}
\caption{Upper left panel: the 2.5-day-bin AP $\gamma$-ray light curve.
		 Lower left panel: the 2D plane contour plot of the WWZ power for the light curve.
		 Lower right panel: the LSP power spectrum for the light-curve (red solid line) and
		                    the time-averaged WWZ power (black solid line).}
\label{PKS_0301-243_ap}
\end{figure*}

We use the two techniques,
the maximum likelihood optimization (ML) and the exposure-weighted aperture photometry \citep[AP;][]{Corbet2007,Kerr2011},
to construct the $\gamma$-ray light curves of PKS 0301$-$243.
The 30-day-bin ML light curve is generated by employing the unbinned maximum likelihood fitting technique. 
In this step,  the Tool \emph{gtlike} is employed for each time-bin,
and the events are selected in a circle ROI of $15^{\circ}$ centered at the coordinates of the target.
We use the same parameter value as that in the new model file for all the sources in the ROI and freeze the spectral parameters except for the target.
The 30-day-bin ML light curve is shown in the left upper panel of Fig. \ref{PKS_0301-243}.
For testing whether the power peaks vary with different length of time-bin,
we also produce the  ML light curve with the 10-day-bin, which is shown in the left upper panel of Fig. \ref{PKS_0301-243_10d}.
The light curve can also be produced by the method of the exposure-weighted aperture photometry.
In this method, we calculate the probabilities for each photon with the \emph{Fermi}-Tool \emph{gtsrcprob}, and then
sum the probabilities of each photon within $1^{\circ}$ radius centered on the position of target for each 2.5-day-bin, in which
the counts are weighted by its relative exposure for each time-bin.
The AP light curve is shown in the left upper panel of Fig. \ref{PKS_0301-243_ap}.

We notice that there is an isolated large flare around MJD 55320. 
To avoid its impact on searching for quasi-periodic variability, we remove this flare in the following quasi-periodicity analyses.

\subsection{Searching for quasi-periodic variability}
\label{subsec:Search qpo}
\subsubsection{Analyses on $\gamma$-ray data}

\begin{figure*}
\centering
\includegraphics[scale=0.5]{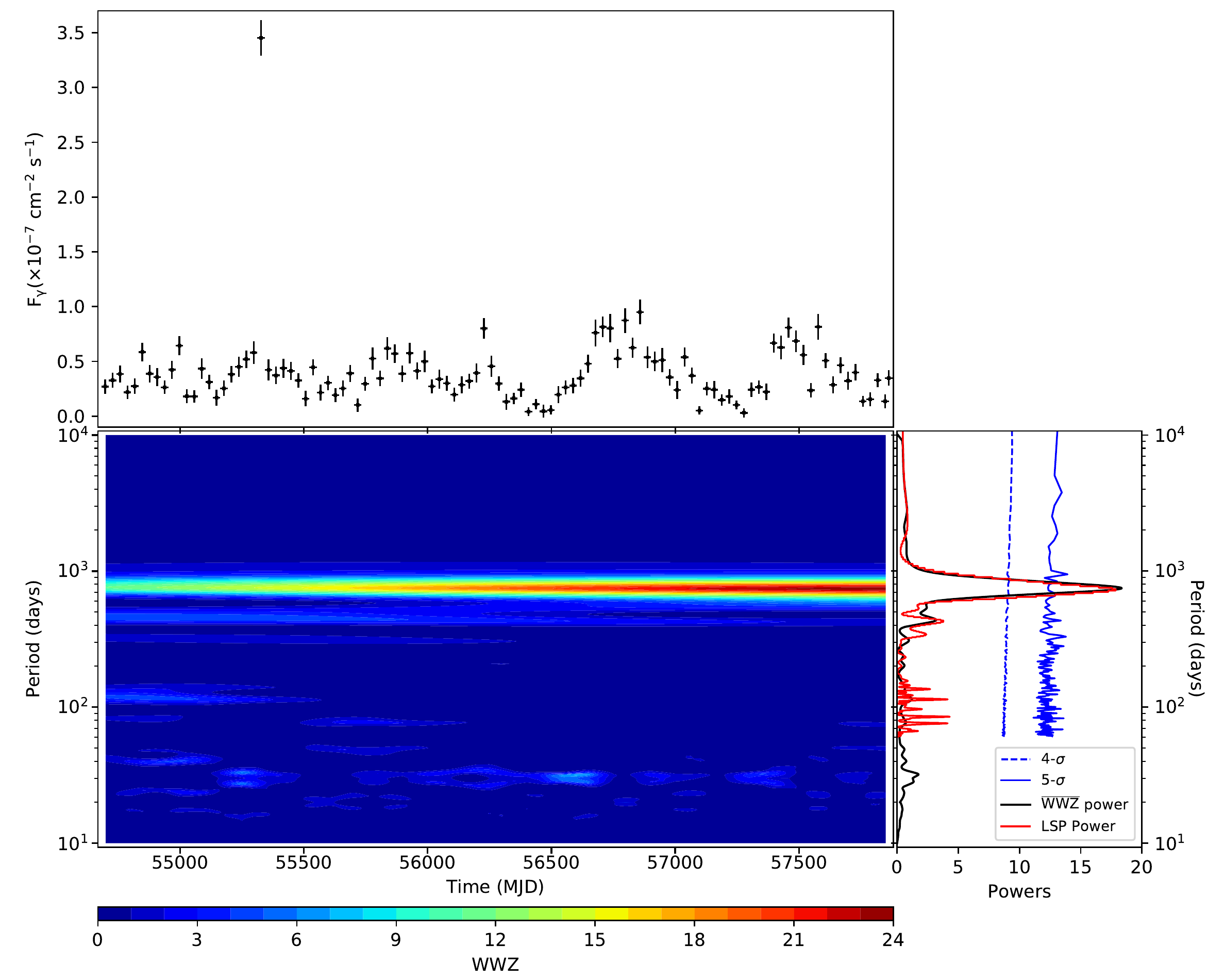}
\caption{Upper left panel: the 30-day-bin ML $\gamma$-ray light-curve.
		 Lower left panel: the 2D plane contour plot of the WWZ power of the light-curve.
		 Lower right panel: the LSP power spectrum for the light-curve (red solid line) and the time-averaged WWZ power (black solid line);
		                               the blue dashed and solid lines represent the 4-$\sigma$ and 5-$\sigma$ confidence level, respectively.}
\label{PKS_0301-243}
\end{figure*}

We apply the two widely used methods, Lomb-Scargle Periodogram
\citep[LSP;][]{Lomb1976,Scargle1982} and Weighted Wavelet Z-transform \citep[WWZ;][]{Foster1996}, to the $\gamma$-ray light curves.
For the 30-day-bin ML light-curve, 
 three power spectra, LSP power, WWZ power and time-averaged WWZ power, are shown in Fig. \ref{PKS_0301-243}.
A strong peak near a period cycle of $2.1\pm0.3$ yr appears, in which 
the maximum power is $>$ 18.6 times of the mean power value.
The probability (Prob) for obtaining a power larger than the maximum power from the noise is $< 1.58\times10^{-9}$
(corresponding to a $> 6.0\sigma$ significance level).
The Prob(P $>$ Pn) is assessed through the formula: $\rm Prob(P > Pn) = (1 - 2\times\frac{Pn}{N-1})^{(N-3)/2}$
with the normalization from \citet{HorneBaliunas1986}, where N=105 is the number of time-bin in the month-bin light curve.
We correct the probability in the range of $1/3000\rm\ day^{-1}$ - $1/60\rm\ day^{-1}$ with the ``trial factor = 50" (the number of sampled independent frequencies)  \citep{Zechmeister2009},
and find that the false-alarm probability (FAP) is less than $7.8\times10^{-8}$, corresponding to $> 5.4\sigma$.
By fitting the power peak with Gaussian-function,
we derive the period cycle of $763.3\pm114.9$ days. The uncertainty of the period is evaluated
based on the half width at half maximum (HWHM) of the Gaussian fitting. 

In order to evaluate the {\it precise} significance of the signal,
we use the method in \citet{Emmanoulopoulos2013} \citep[also see][]{1553,Bhatta2016} to simulate light curves $3\times10^{6}$ times based on the
obtained best-fitting result of power spectral density (with the form of $P(f)\sim1/f^{\alpha} + c$, where $c$ represents the Poisson noise level) and the
probability density function of observed variation.
We then derive the significance curves of 5-$\sigma$ and 4-$\sigma$ based on the simulations,
which are shown in the lower right panel of Fig. \ref{PKS_0301-243}.
The significance of the signal is $\simeq5.4\ \sigma$.
We also calculate the power spectra of the 10-day ML light curve and 2.5-day-bin AP light-curve,
which are shown in Fig. \ref{PKS_0301-243_10d} and Fig. \ref{PKS_0301-243_ap}, respectively.
In these two power spectra, we also find strong signals at $\sim$2.1 yr.

\begin{figure*}
\centering
\includegraphics[scale=0.4]{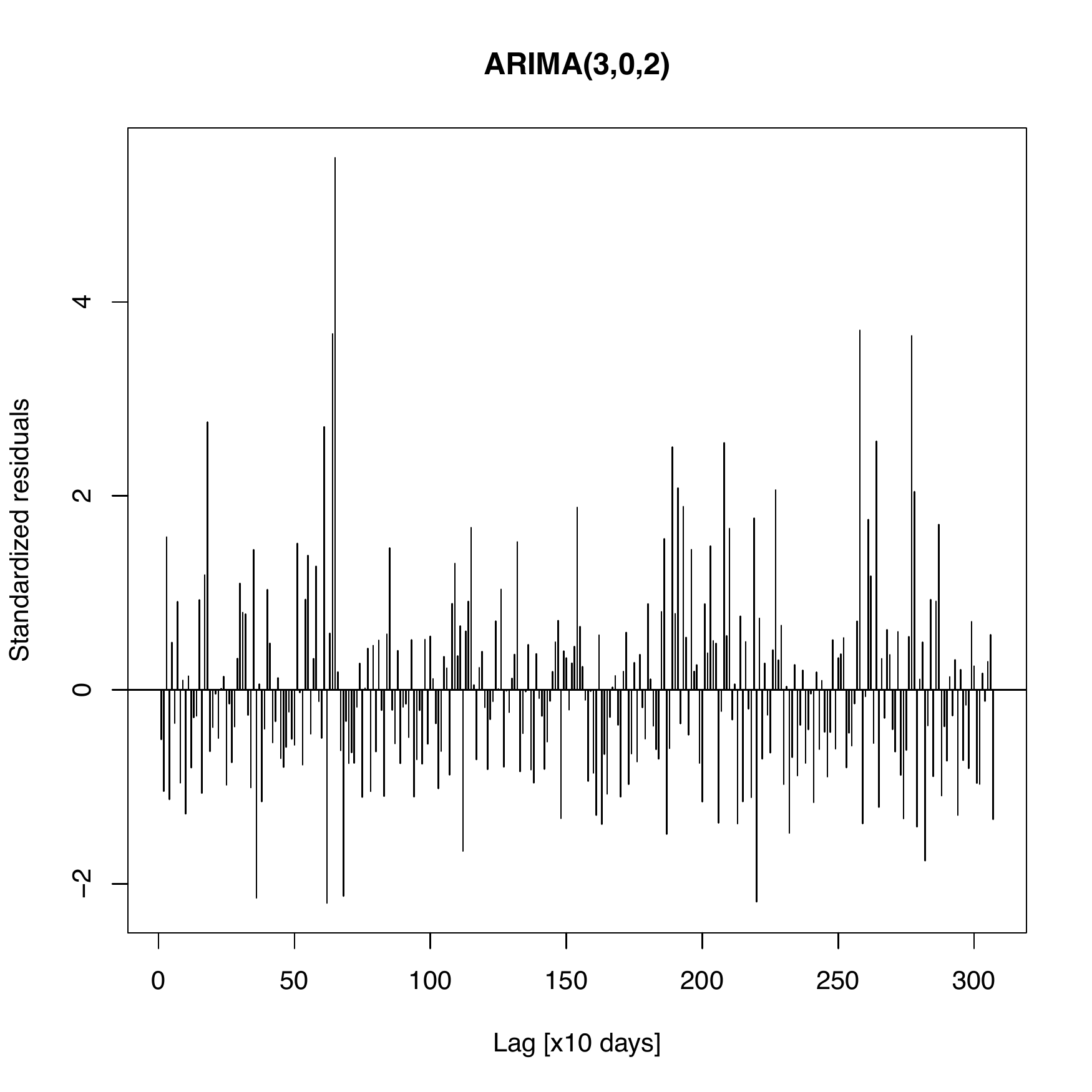}
\includegraphics[scale=0.4]{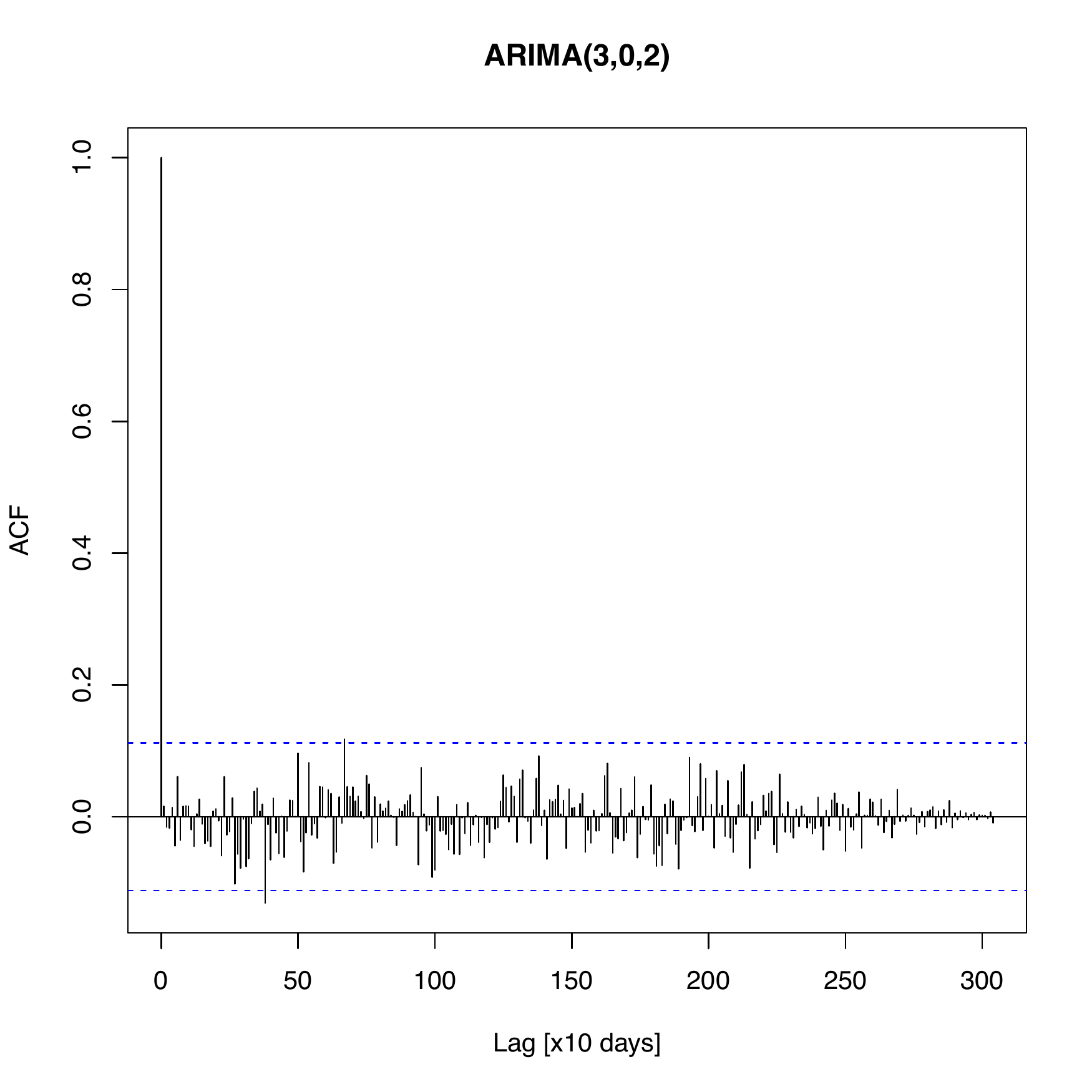}
\caption{Results of ARIMA(3,0,2) model fitting the 10-day-bin $\gamma$-ray light curve. {\it Left: }
standard residuals from the fitting; {\it right: } ACF of the residuals. The
dashed horizontal lines represent the 95\% confidence level.}
\label{arima_image}
\end{figure*}

In order to further check the reliability of the quasi-periodic signal, 
we fit $\gamma$-ray light curve with autoregressive integrated moving average (ARIMA) models \citep{box,Hamilton1994,Chatfield2003} to 
assess whether the signal is consistent with a stochastic origin of autoregressive noise.
We use the Akaike Information Criterion \citep[AIC;][]{Akaike1973} to select the best-fit model.
In Table~\ref{arima_table}, we show the AIC values for 72 ARIMA models fitting the 10-day-bin $\gamma$-ray light curve.
One can see that the ARIMA (3,0,2) model, with the minimum AIC value of 1355\footnote{We note that the AIC values of several 
models [e.g., ARIMA(1,0,0), ARIMA(1,0,2), and ARIMA(1,0,3)] are very close to 1355. This indicates that more data are needed to clearly distinguish these models.}, is the best-fit one.
In Fig.~\ref{arima_image},  we show the standard residuals and the auto-correlation function (ACF) of the residuals for the best-fit model.
It can be seen that there is a spike at the lag of 660 days that exceeds the 95\% confidence limit.
This marginal evidence indicates that the $\gamma$-ray quasi-periodic variability may not be produced by such type of stochastic processes.

We fold the events within a square region of interest (ROI)
of $20^\circ\times20^\circ$ centered at the position of PKS 0301$-$243 into 15 uniform bins based on orbital phase with
the phase zero corresponding to MJD 54,682.66. We then fit the data in each phase bin by using the above best-fitting model-file 
to obtain the phase-resolved likelihood results. In Fig. \ref{PKS_0301-243_phase}, one can see that this folded light curve varies with the phase, 
indicating substantial variability in the source brightness (see the upper panel of Fig. \ref{PKS_0301-243_phase});
but no variability appears in its spectral shape (see the lower panel of Fig. \ref{PKS_0301-243_phase}).

\subsubsection{Analyses on optical and X-ray data}
\label{subsec:xRay}

\begin{figure*}
\centering
\includegraphics[scale=0.6]{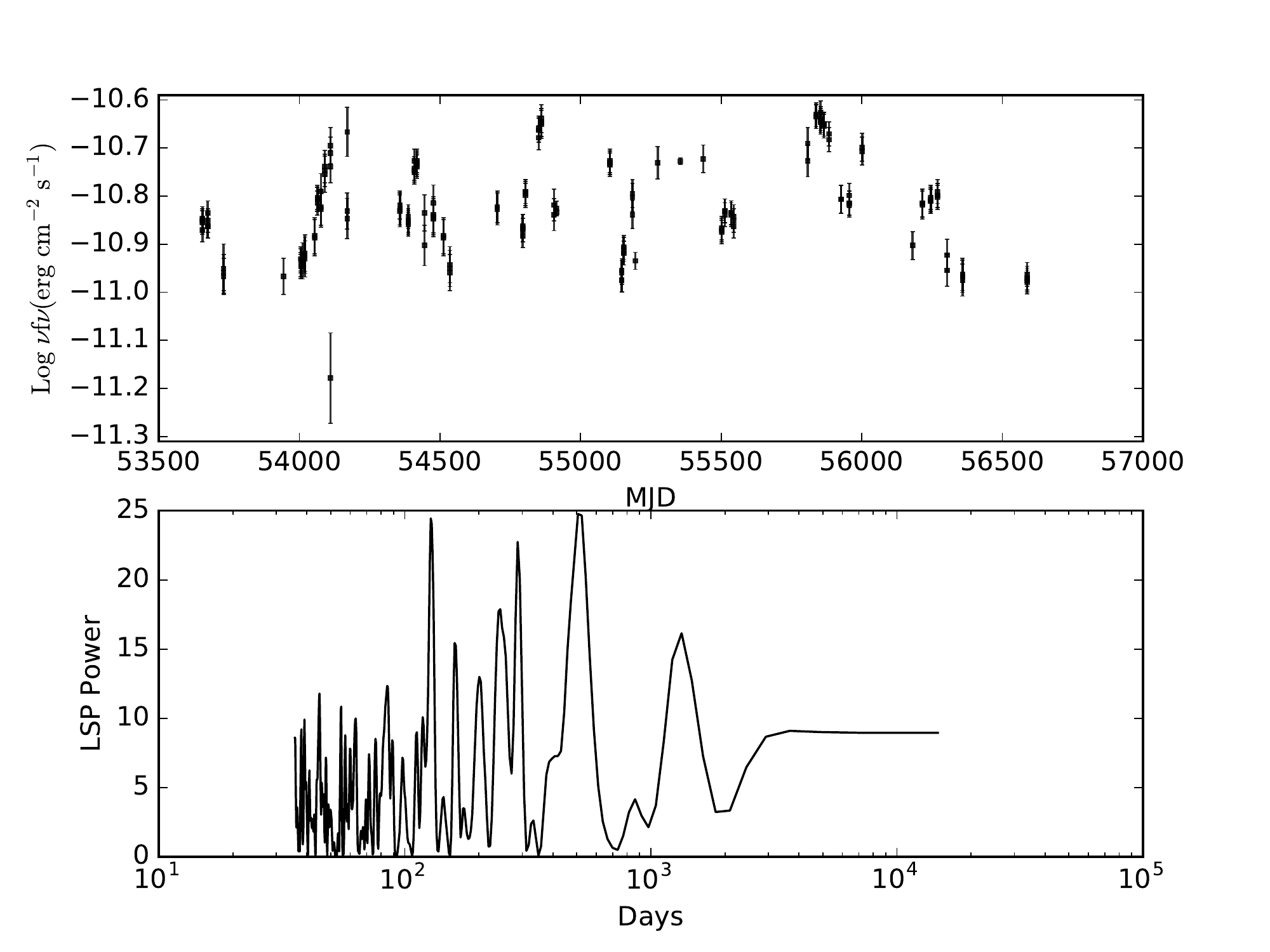}
\caption{Upper panel: the optical light curve obtained in the Catalina Sky Surveys (data from the ASI Science Data Center).
		 Lower panel: the LSP power spectrum for the optical light curve.}
\label{optical}
\end{figure*}

\begin{figure*}
\centering
\includegraphics[scale=0.6]{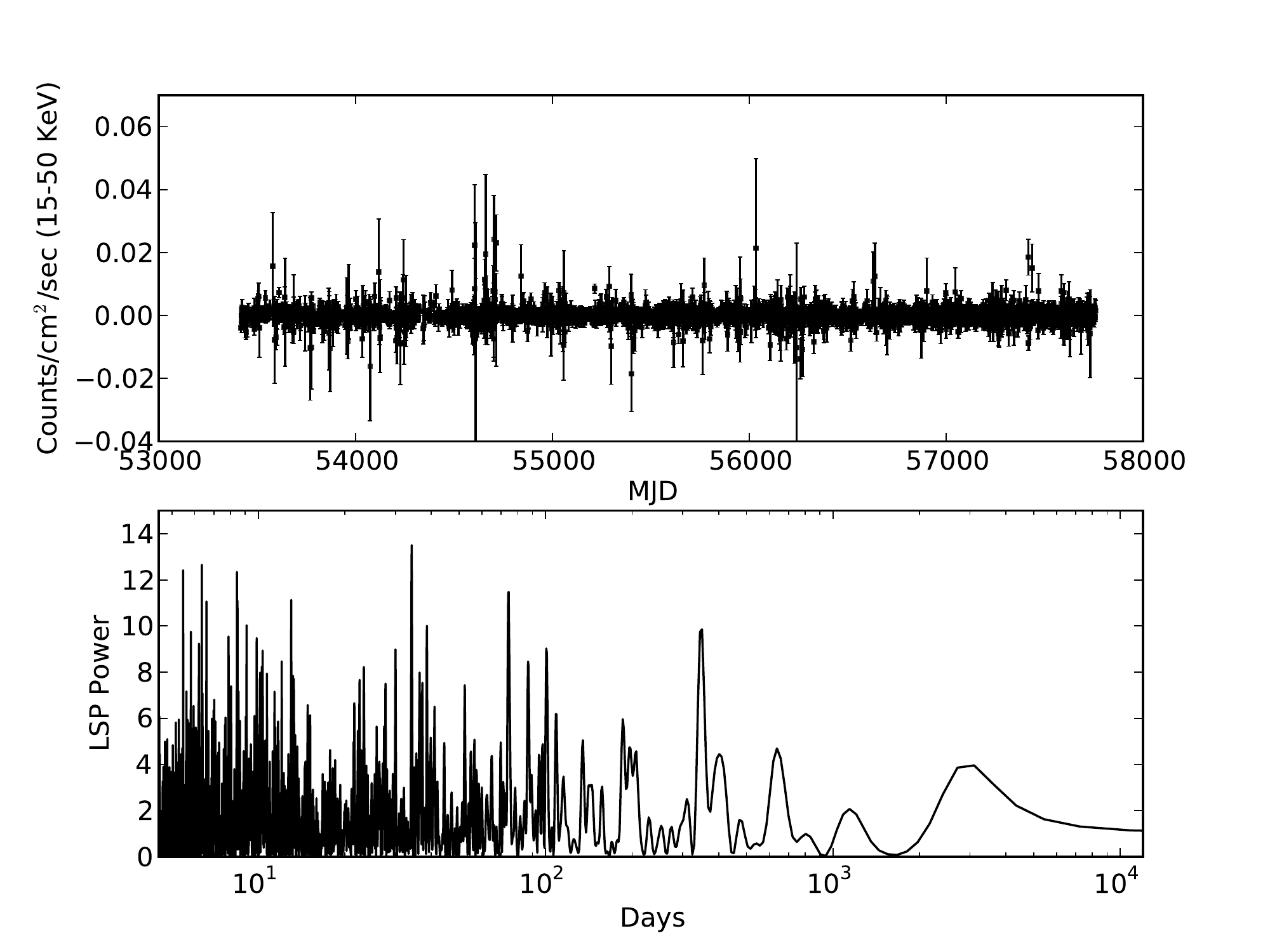}
\caption{Upper panel: the Swift-BAT X-ray light curve  (data from the ASI Science Data Center). Lower panel: the LSP power spectrum for the light curve.}
\label{x-ray}
\end{figure*}

We also search for quasi-periodic signal in the optical and X-ray data from this source. The long-term 
optical data from the Catalina Sky Surveys covering from 2005 October to 2013 October
and daily averaged X-ray data from Swift-BAT covering from 2005 February to 2017 January
are shown in the upper panels of Fig. \ref{optical} and Fig. \ref{x-ray}, respectively.
The LSP powers of the optical data and X-ray data  are respectively shown in the lower panels of Fig. \ref{optical} and Fig. \ref{x-ray}. 
No obvious peak is found in the corresponding powers.
It is noted that the X-ray data are weakly variable.

\section{SUMMARY AND DISCUSSION}
\label{sec:summary}

Possible $\gamma$-ray QPOs have been reported in several blazars \citep[e.g.,][]{Sandrinelli2014,1553,Sandrinelli2016a,Sandrinelli2016b,Sandrinelli2017,Zhang2017a,Zhang2017b}.
However, the significance of the claimed QPOs is not very high.
In this paper, we report the first detection of $\gamma$-ray quasi-periodic modulation at a nominal confidence level of \textbf {$\sim5\sigma$} in PKS 0301$-$243.
No quasi-periodic modulation is found in its optical and X-ray data.

In PG 1553$+$113, the quasi-periodic variabilities in optical and $\gamma$-ray data have the same period cycle \citep{1553}.
In PKS 2155$-$304, the periods of optical and $\gamma$-ray quasi-periodic variabilities are different \citep{Sandrinelli2014}.
In PKS 0426$-$380, no optical quasi-periodic variability is found \citep{Zhang2017a}.
The lack of optical and X-ray quasi-periodic variabilities may be because of the optical and X-ray originating 
from the different region that does not contribute $\gamma$-rays.
If the lack of optical and X-ray quasi-periodic variabilities is confirmed by 
futuer long-term monitoring, it would challenge the most popular one-zone blazar
emission model in which optical, X-ray and $\gamma$-ray emissions are assumed to be produced in the same region \citep{Zhang2017a}.

The mechanism causing the $\gamma$-ray quasi-periodic modulation in blazars is poorly understood.
Given that the $\gamma$-rays are produced in the jet,
two possibilities may account for the $\gamma$-ray quasi-periodic variabilities in blazars \citep[e.g.,][]{1553}:
(i) pulsational accretion flow instabilities may induce a quasi-periodic injection of plasma into the  jet,
hence a quasi-periodic modulation appears in the $\gamma$-ray flux from the jet; and
(ii) the Doppler magnification factor changes periodically caused by jet precession/rotation.

\begin{figure*}
\centering
\includegraphics[scale=0.7]{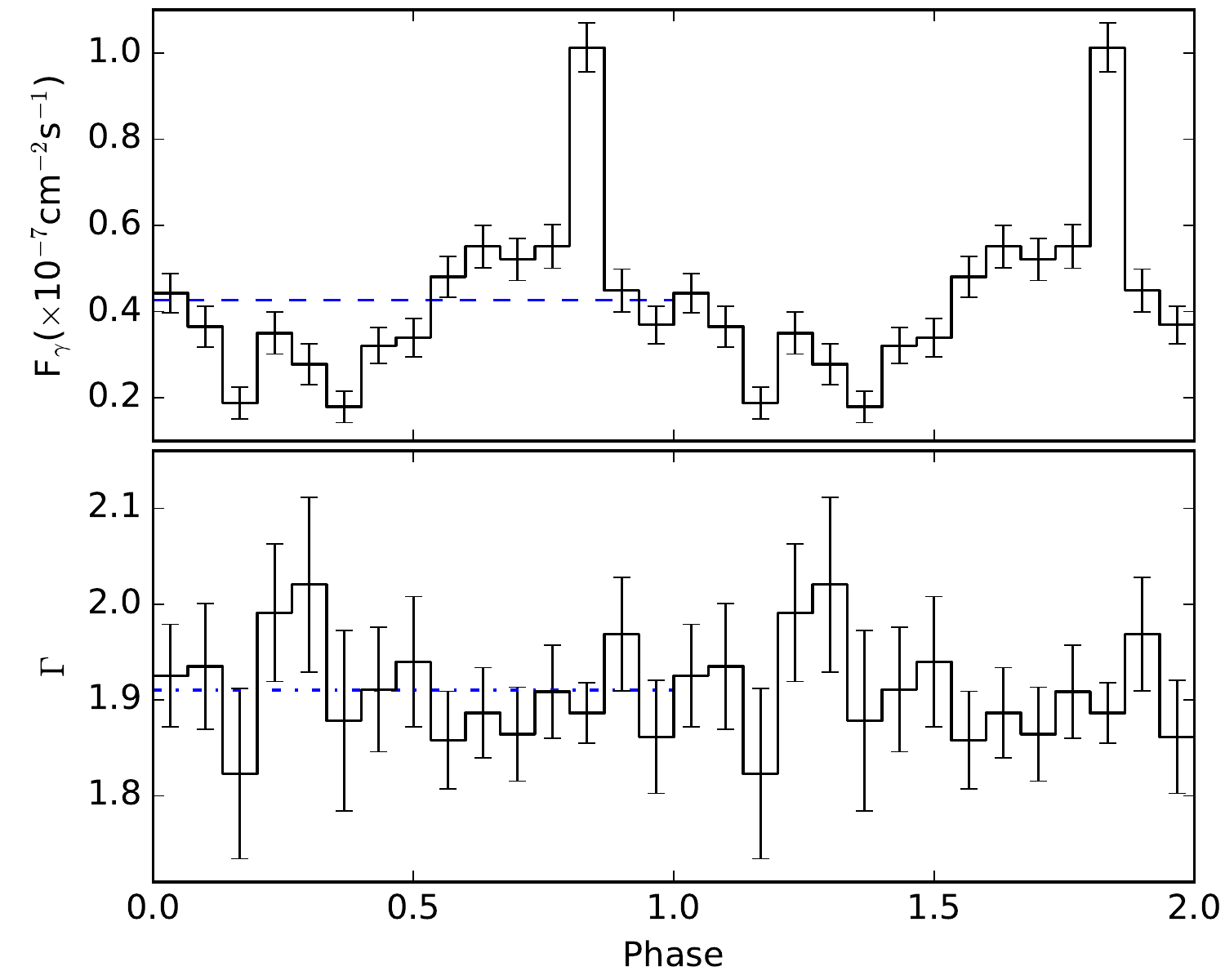}
\caption{Upper panel: the epoch-folded pulse shape above 0.1 GeV with the period of 763.3-day.
                      The blue dashed line is the mean flux. For clarity, we show two period cycles.
             Lower panel: the gamma-ray spectral index ($\Gamma$) in the phase. The blue dashed-dotted line is the mean value of $\Gamma$.}
\label{PKS_0301-243_phase}
\end{figure*}

Note that in our case the $\gamma$-ray photon index does not vary with the phase  (see the lower panel of Fig. \ref{PKS_0301-243_phase}).
The gamma-ray photon index is mainly determined by the high-energy electrons distribution.
    For HSP, the electron cooling is inefficient  \citep[e.g.,][]{Ghisellini08,Yan14}, and the electron distribution is mainly governed by the acceleration mechanism in the jet.
This result indicates that the process yielding the QPO in the $\gamma$-ray flux would not have an impact on the acceleration process.
The first possible origin for the QPO outlined above would have an impact on the energy outflow efficiency which is relative to the acceleration process in the jet \citep[e.g.,][]{1553}.
Therefore, our results may prefer to the second origin, i.e., jet precession.
The jet precession could be the result of a helical jet \citep[e.g,][]{Rieger04,Ko2016}.
Furthermore, a binary SMBH system would be involved in the formation of a helical jet \citep[e.g,][]{Ko2016,Sobacchi}.
Within such a scenario, the observed 2.1 yr period is the orbital time, and the equivalent intrinsic orbital time $P_{\rm int}$=$P_{\rm obs}/(1+z)$.
The central SMBH of PKS 0301$-$243 is $\sim8\times10^8\ M_{\odot}$ \citep{Ghisellini10}.
Assuming the total mass of the binary SMBH of $10^9\ M_{\odot}$,  
the binary system size would be $\sim0.006\ $pc. At this stage, gravitational wave
emission may be non-negligible in carrying away the energy.

In the jet precession model, the issue of the lack of optical and X-ray quasi-periodic variabilities could be resolved if the optical
and X-ray radiations originate from a large region where the Doppler boosting is weak.
Systematic sample study on QPOs at different electromagnetic frequencies in blazars may reveal deep physics of the jet \citep[e.g.,][]{Sandrinelli2016a}.

The $\gamma$-ray QPO in PKS 0301$-$243 is the first detection of such a kind of signal in blazars at a confidence level of $\sim5\sigma$.
Since there were barely four nominal quasi-periods in the currently {\it Fermi}-LAT data, this result certainly requires confirmation.
Fortunately, our claim for a QPO should be tested rather soon, as the next flux maximum would be expected in 2018.

\begin{figure*}
\centering
\includegraphics[scale=0.6]{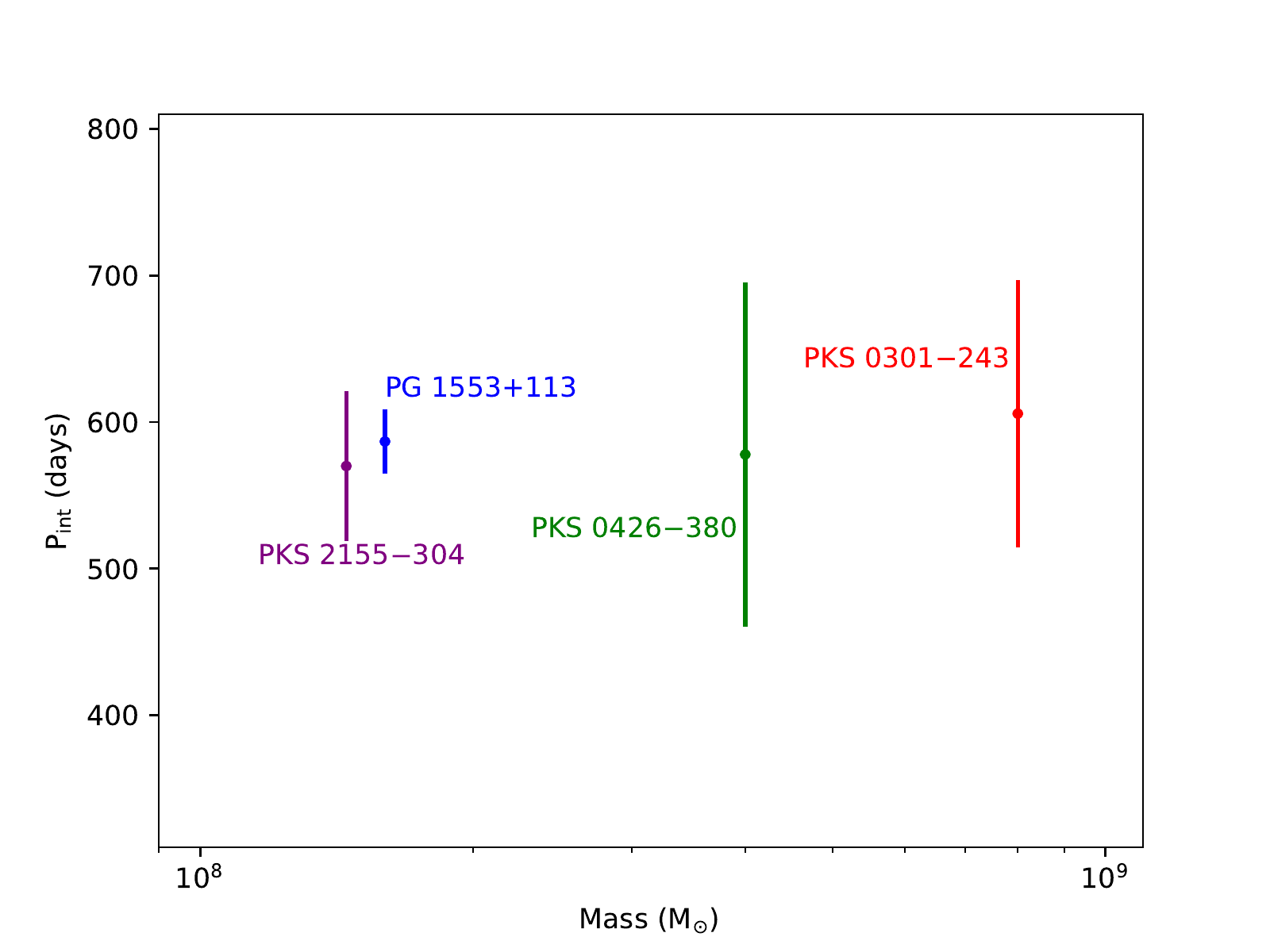}
\caption{The relation between $P_{\rm int}=P_{\rm obs}/(1+z)$ and SMBH mass. The $\gamma$-ray intrinsic periods
                       for PKS 2155$-$304, PG 1553$+$113 and PKS 0426$-$380 are from \citet{Zhang2017a}, \citet{1553},  and \citet{Zhang2017b}, respectively.
                      The SMBH mass for PKS 2155$-$304, PG 1553$+$113, PKS 0426$-$380 and PKS 0301$-$243 
                      are from \citet{Zhang2005}, \citet{Ghisellini2014}, \citet{Sbarrato2012} and \citet{Ghisellini10}, respectively.}
\label{pvsm}
\end{figure*}

Finally, we would like to stress these claimed $\gamma$-ray QPOs in blazars are different from the X-ray QPO in BH X-ray binaries and narrow-line Seyfert 1
galaxy \citep{Zhang2017b}. For the X-ray QPOs, there is an inverse linear relation between QPO frequency
and BH mass \citep[e.g.,][]{Abramowicz2004,Torok2005,Re06,Pan2016}. This relation spans from stellar-mass to SMBH.
No such relation is found in $\gamma$-ray QPO in blazars (Fig.~\ref{pvsm}). 
It seems that the intrinsic period of $\gamma$-ray QPO in blazars is independent on the SMBH mass.
Moreover, the relation of the $\gamma$-ray QPO frequency-BH mass significantly deviates from the inverse relation found in the X-ray QPOs.
The X-ray and $\gamma$-ray QPOs provide us different insights into the BH - jet system.
$\\$

We thank the anonymous referees for useful and constructivecomments.
Part of this work is based on archival data, software or online services provided by the ASI Science Data Center (ASDC).
We acknowledge the financial support from the 973 Program of China under grant 2013CB837000,
the National Natural Science Foundation of China (NSFC-11433004, NSFC-11525313, NSFC-11573060, NSFC-11573026 and NSFC-11661161010),
and the Key Laboratory of Astroparticle Physics of Yunnan Province (No. 2016DG006).
D.-H. Yan is very grateful to Prof. Yefei Yuan (USTC) for helpful discussions.
The work of D.-H. Yan is supported by the CAS ``Light of West China" Program.


\bibliography{ApJS}

\clearpage
\begin{table*}
\begin{center}
\caption{AIC values for the ARIMA models fitting the 10-day-bin light curve.}
\begin{tabular}{cc|c|c|c|c|c|c|c|c|c|c|c|ccccc}
\hline\hline
\multicolumn{2}{c|}{\multirow{3}{*}{ARIMA({\it p,d,q})}} & \multicolumn{12}{c}{MA({\it q})} \\ \cline{3-14}
                              & &\multicolumn{6}{|c}{{\it d}=0} & \multicolumn{6}{|c}{{\it d}=1}\\ \cline{3-14}
                              && MA(0) & MA(1) & MA(2) & MA(3) & MA(4) & MA(5) &MA(0) & MA(1) & MA(2) & MA(3) & MA(4) & MA(5) \\\cline{2-14}
\hline
\multicolumn{1}{c|}{\multirow{6}{*}{AR({\it p})}}   &    AR(0)  & 1475    & 1388   & 1370  & 1368 & 1366 & 1368    &  1425 & 1391 & 1365 & 1363 & 1365 & 1367   \\\cline{2-14}
\multicolumn{1}{c|}{\multirow{6}{*}{}}                   &    AR(1)  &1359    & 1361     & 1359  & 1359 &  1361 & 1363 &1410 & 1363    &1364    & 1365  &1366  & 1368   \\\cline{2-14}
\multicolumn{1}{c|}{\multirow{6}{*}{}}                   &    AR(2)  & 1361   & 1363   &  1360   & 1360 & 1362 &  1364 &1397 & 1364   & 1363    &1366   & 1360 & 1369   \\\cline{2-14}
\multicolumn{1}{c|}{\multirow{6}{*}{}}                   &    AR(3)  & 1362   &1360    & {\bf 1355}  &  1362  & 1364 & 1364  &1387 & 1365    &1366    & 1368 & 1370 & 1371    \\\cline{2-14}
\multicolumn{1}{c|}{\multirow{6}{*}{}}                   &    AR(4)  &  1362  &1363   & 1362   & 1364  & 1365 & 1366 &1383& 1367     & 1368   & 1370  & 1360 & 1371    \\\cline{2-14}
\multicolumn{1}{c|}{\multirow{6}{*}{}}                   &    AR(5)  &  1363  & 1362   & 1364    & 1356  & 1360 & 1368 & 1373&  1368   &  1368   & 1370  & 1371 & 1373   \\\cline{2-14}
\hline\hline
\end{tabular}
\label{arima_table}
\end{center}
\end{table*}
\end{document}